\documentclass[aps,prb,superscriptaddress,twocolumn,showpacs,intlimits,amsmath,amssymb,floats,longbibliography]{revtex4-1}

\usepackage{times}
\usepackage{epsfig}
\usepackage{graphicx}
\usepackage{pstricks}
\usepackage{dcolumn}		% Align table columns on decimal point
\usepackage{bm}			% bold math
\usepackage{ulem}
\usepackage{units}		% Formatting number and unit
\usepackage{color}
\usepackage[english]{babel}
\usepackage[permil]{overpic}
\usepackage{float}
\usepackage[colorlinks=true,allcolors=blue]{hyperref}  %hyperref still needs to be put at the end!

\begin{document}

\title{On the nature of valence charge and spin excitations via multi-orbital Hubbard models for infinite-layer nickelates}

\author{Emily M. Been}
\affiliation{Stanford Institute for Materials and Energy Sciences, SLAC National Accelerator Laboratory, 2575 Sand Hill Road, Menlo Park, California 94025, USA}
\affiliation{Department of Physics, Stanford University, Stanford, California 94305, USA}
\author{Kuan H. Hsu}
\affiliation{Stanford Institute for Materials and Energy Sciences, SLAC National Accelerator Laboratory, 2575 Sand Hill Road, Menlo Park, California 94025, USA}
\affiliation{Department of Materials Science and Engineering, Stanford University, Stanford, California 94305, USA}
\author{Yi Hu}
\affiliation{Stanford Institute for Materials and Energy Sciences, SLAC National Accelerator Laboratory, 2575 Sand Hill Road, Menlo Park, California 94025, USA}
\affiliation{Department of Materials Science and Engineering, Stanford University, Stanford, California 94305, USA}
\author{Brian Moritz}
\affiliation{Stanford Institute for Materials and Energy Sciences, SLAC National Accelerator Laboratory, 2575 Sand Hill Road, Menlo Park, California 94025, USA}
\author{Yi Cui}
\affiliation{Department of Materials Science and Engineering, Stanford University, Stanford, California 94305, USA}
\author{Chunjing Jia}
\affiliation{Stanford Institute for Materials and Energy Sciences, SLAC National Accelerator Laboratory, 2575 Sand Hill Road, Menlo Park, California 94025, USA}
\author{Thomas P. Devereaux}
\affiliation{Stanford Institute for Materials and Energy Sciences, SLAC National Accelerator Laboratory, 2575 Sand Hill Road, Menlo Park, California 94025, USA}
\affiliation{Department of Materials Science and Engineering, Stanford University, Stanford, California 94305, USA}
\email{tpd@stanford.edu}

\date{\today}% It is always \today, today,
             %  but any date may be explicitly specified

\begin{abstract}
%%% Leave the Abstract empty if your article does not require one, please see the Summary Table for full details.
Building upon the recent progress on the intriguing underlying physics for the newly discovered infinite-layer nickelates, in this article we review an examination of valence charge and spin excitations via multi-orbital Hubbard models as way to determine the fundamental building blocks for Hamiltonians that can describe the low energy properties of infinite-layer nickelates. We summarize key results from density-functional approaches, and apply them to the study of x-ray absorption to determine the valence ground states of infinite-layer nickelates in their parent form, and show that a fundamental $d^9$ configuration as in the cuprates is incompatible with a self-doped ground state having holes in both $d_{x^2-y^2}$ and a rare-earth-derived axial orbital. When doped, we determine that the rare-earth-derived orbitals empty and additional holes form low spin $(S=0)$ $d^8$ Ni states, which can be well-described as a doped single-band Hubbard model. Using exact diagonalization for a 2-orbital model involving Ni and rare earth orbitals, we find clear magnons at 1/2 filling that persist when doped, albeit with larger damping, and with a dependence on the precise orbital energy separation between the Ni- and rare-earth-derived orbitals. Taken together, a full two-band model for infinite-layer nickelates can well describe the valence charge and spin excitations observed experimentally.

%For full guidelines regarding your manuscript please refer to \href{http://www.frontiersin.org/about/AuthorGuidelines}{Author Guidelines}.

%As a primary goal, the abstract should render the general significance and conceptual advance of the work clearly accessible to a broad readership. References should not be cited in the abstract. Leave the Abstract empty if your article does not require one, please see \href{http://www.frontiersin.org/about/AuthorGuidelines#SummaryTable}{Summary Table} for details according to article type. 

\end{abstract}

\maketitle

%\section{Article types}

%For requirements for a specific article type please refer to the Article Types on any Frontiers journal page. Please also refer to  \href{http://home.frontiersin.org/about/author-guidelines#Sections}{Author Guidelines} for further information on how to organize your manuscript in the required sections or their equivalents for your field

% For Original Research articles, please note that the Material and Methods section can be placed in any of the following ways: before Results, before Discussion or after Discussion.

%Original Research
%Original Research articles report on primary and unpublished studies. Original Research may also encompass confirming studies and disconfirming results which allow hypothesis elimination, reformulation and/or report on the non-reproducibility of previously published results. Original Research articles are peer-reviewed, have a maximum word count of 12,000 and may contain no more than 15 Figures/Tables. Authors are required to pay a fee (A-type article) to publish an Original Research article. Original Research articles should have the following format: 1) Abstract, 2) Introduction, 3) Materials and Methods, 4) Results, 5) Discussion.

%1.15. Sections
%The manuscript is organized by headings and subheadings. The section headings should be those appropriate for your field and the research itself. You may insert up to 5 heading levels into your manuscript (i.e.,: 3.2.2.1.2 Heading Title).

%For Original Research articles, it is recommended to organize your manuscript in the following sections or their equivalents for your field:

\section{Introduction}

%Succinct, with no subheadings.
The discovery of experimental superconductivity in infinite-layer nickelates \citep{Denver_nickelateDiscovery_2019},  20 years after they were theoretically proposed \citep{Anisimov_1999}, has unveiled new territory to probe the unknowns of unconventional superconductivity. The comparison between the superconducting nickelates to the cuprates has proven to be a rich area of inquiry. Significant progress has been made in the last two years, but many mysteries about this novel superconducting family remain unsolved. As pointed out early on \citep{Pickett_2004}, the bandstructure for LaNiO$_2$ while being primarily $3d^9$ Ni near the Fermi level has additional small Fermi pockets of largely axial character involving Ni $3d_{z^2}$ and La $5d_{z^2}$ orbitals. Including a Hubbard $U$ on nickel splits the $d^9$ states but still leaves itinerant states at the Fermi level, indicating that the parent compound for infinite-layer nickelate is not analogous to undoped CuO$_2$ as an antiferromagnetic Mott insulator. Moreover, the location of the centroid of the oxygen states in LaNiO$_2$ lies at lower energies than CuO$_2$ compounds, moving them closer to the boundary between Mott-Hubbard systems and charge-transfer insulators in the Zaanen-Sawatzky-Allen scheme \citep{ZSA}. While there is scant evidence that LaNiO$_2$ is antiferromagnetic, resonant inelastic x-ray scattering (RIXS) has clearly revealed propagating magnons, yielding a spin exchange energy $\sim$ 60 meV in the parent compound \citep{Lu_2021}. These excitations persist when doped, albeit with larger damping \citep{Lu_2021}, in a way that is not too disimilar to paramagnon excitations in doped cuprates \citep{Jia2014,LeTacon_2011,Dean_2013}.

Thus there already is a rich amount of information from which an estimate for a fundamental low energy model Hamiltonian can be obtained and used to model the phase diagram of infinite-layer nickelates. In this article we review efforts to combine density functional approaches with cluster exact diagonalization to determine such a model. Specifically we utilize x-ray absorption (XAS) as a tool to determine valence charge states in the undoped and doped infinite-layer nickelates, arriving at a low-energy two orbital model Hamiltonian. We then perform calculations on finite-sized clusters to determine the dynamic spin structure factor in undoped and doped systems. Our results yield a consistent description of charge valence and spin excitation states in the infinite-layer nickelates, indicating that a two-orbital model likely contain the fundamental ingredients needed to explore superconductivity and the rich phase diagram in these compounds.

\section{Methods} %MATERIALS AND METHODS

%This section may be divided by subheadings and should contain sufficient detail so that when read in conjunction with cited references, all procedures can be repeated. 

\subsection{DFT}
\label{sec:DFT}
\begin{figure*}
    \centering
    \includegraphics[width=2\columnwidth]{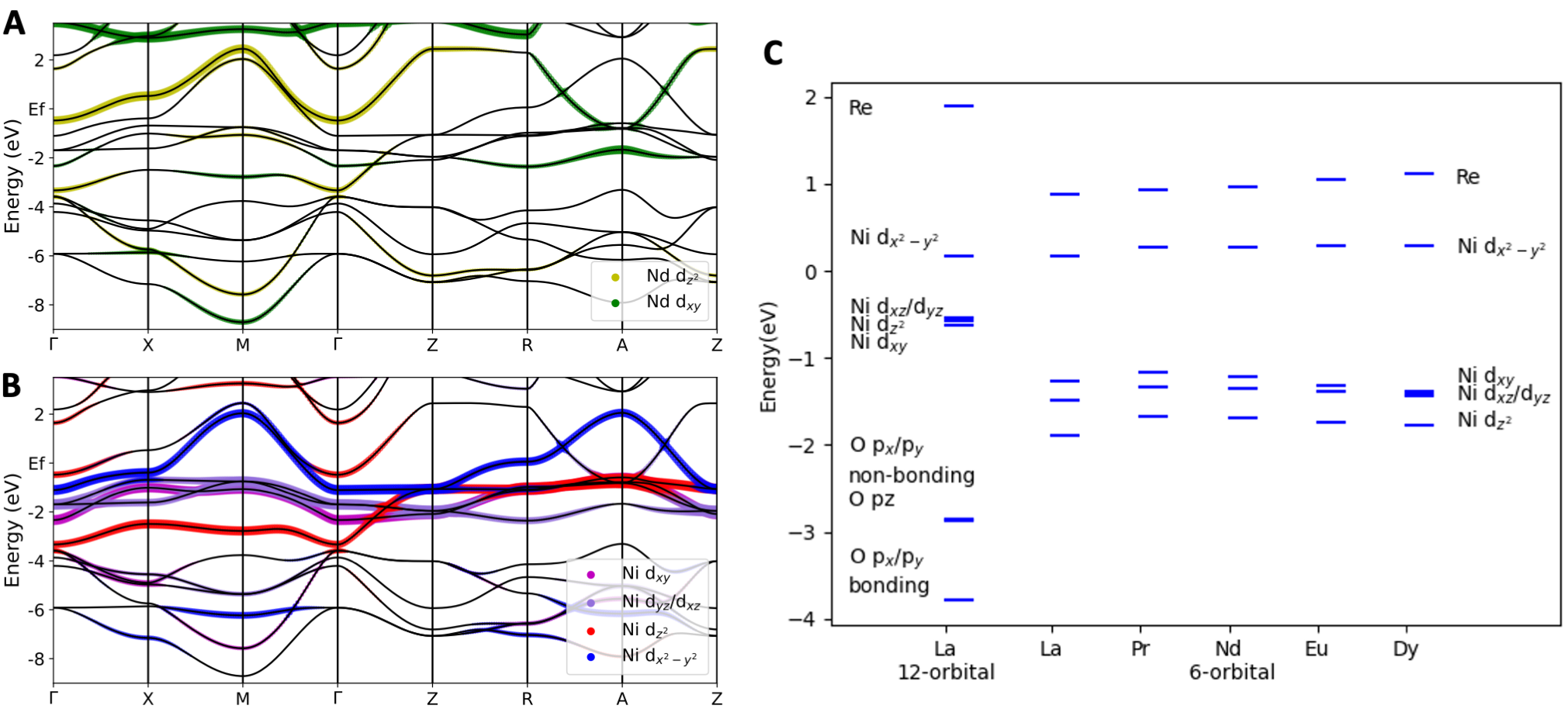}
    \caption{(A) DFT calculated band structure and projected Nd $5d$ orbital content for infinite-layer nickelate NdNiO$_2$. (B) DFT calculated band structure and projected Ni $3d$ orbital content for NdNiO$_2$. (C) Atomic energy level diagram for infinite-layer nickelates ReNiO$_2$, where Re = La, Pr, Nd, Eu and Dy. On-site energies are obtained by 6-orbital and 12-orbital Wannier downfolding from DFT calculated band structures. All energy levels are relative to the Fermi level of the corresponding material.}
    \label{fig:BS_ED}
\end{figure*}

The electronic structure of the infinite-layer nickelates ReNiO$_2$ have been evaluated using density functional theory (DFT) in the generalized gradient approximation (GGA) for the exchange-correlation functional as implemented in Quantum Espresso (\cite{QE, PBE}). The band structure near the Fermi energy that was obtained for NdNiO$_2$ is shown in Figs. \ref{fig:BS_ED}{\bf A} and {\bf B}. 

To obtain a microscopic Hamiltonian and understand the atomic energy levels, we downfolded the bandstructure as implemented in Wannier90 (\cite{Wannier90}). Two types of Wannier downfolding have been calculated: (1) a 12-orbital model downfolding, which includes five Ni $3d$ orbitals, one Re $5d$ orbital and six O $2p$ orbitals; and (2) a 6-orbital model downfolding, which includes five Ni $3d$ orbitals and one Re $5d$ orbital. For NdNiO$_2$, the bandstructure corresponding to the 6-orbital model spans an energy range from $\sim$ -3.5eV to $\sim$ 2eV, covering the most prominent low energy features. The bands below $\sim$ -3.5eV for NdNiO$_2$ have predominantly oxygen orbital content. Fig. \ref{fig:BS_ED}{\bf C} summarizes the atomic energy level diagram, which shows the on-site energies from the Wannier downfolding for a series of ReNiO$_2$ materials (Re = La, Pr, Nd, Eu or Dy).

\subsection{X-Ray Absorption from Multi-Orbital Hubbard Hamiltonians} \label{sec:XAS}

We focus on the Ni $L$-edge ($2p \rightarrow 3d)$ XAS utilizing a cluster model for monovalent NiO$_2$ containing 5 Ni $3d$ orbitals and 6 O $2p$ orbitals in a square-planar geometry, and 3 Ni $2p$ orbitals per unit cell in the atomic core, with hybridization parameters obtained from Wannier downfolding as described in the previous section. We evaluate the XAS $\kappa$ for the absorption of a photon having momentum and polarization ${\bf k_i, e_i}$, respectively, as
\begin{equation}
\kappa_{\bf e_i, k_i}(\omega)= \frac{1}{\pi Z} \sum_{i,\nu} e^{-\beta E_i} \Big| \langle\nu| \hat D_{\bf k_i}({\bf e_i})| i \rangle\Big|^2 \delta(\omega-(E_\nu-E_i)),
\label{Eq:XAS}
\end{equation}
Here, $Z$ is the partition function, $E_{\{i,\nu\}}, | \{i,\nu\}\rangle$ are the energies and eigenstates of the initial (ground state) and XAS final states, respectively, obtained from exact diagonalization of the multi-orbital Hubbard model in the cluster Hilbert space. The multi-orbital Hubbard model can be compactly written as
\begin{equation}
\begin{split}
    H=&\sum_{i,j,\sigma} \sum_{\mu,\nu} t_{i,j}^{\mu,\nu} d_{i,\mu,\sigma}^\dagger d_{j,\nu,\sigma} \\
    &+ \frac{1}{2} \sum_i \sum_{\mu,\nu,\mu',\nu'} \sum_{\sigma,\sigma'}U_{\mu,\nu,\mu',\nu'} d_{i,\mu,\sigma}^\dagger d_{i,\nu,\sigma'}^\dagger d_{i,\mu',\sigma'} d_{i,\nu',\sigma}.
    \end{split}
\end{equation}
Here $i,j$ denote sites, $\sigma,\sigma'$ denote spin, and $\nu,\mu,\dots$ denote orbital indices. The operators $d_{i,\mu,\sigma}^\dagger, d_{i,\mu,\sigma}$ create, annihilate holes having spin $\sigma$ in orbital $\mu$ at site $i$, and can represent valence, ligand, or core holes.
The Coulomb matrix elements $U$ are written in terms of Slater-Condon parameters pertaining to the valence hole and core hole states, which can be determined via a variety of methods or simply used as fitting parameters to x-ray spectra. Lastly $\hat D$ is the dipole operator connecting Ni $2p$ core levels to Ni $3d$ states. Addittional details about XAS can be found in various textbooks, {\it i.e.} \citep{deGroot_book}. As the Hilbert space scales exponentially with the number of unit cells, this technique is amenable to only small clusters. However, since XAS is local probe of electronic structure, our computational approach is well suited to determine the effective low energy valence and spin states of Ni, as is well known in many other contexts.

\subsection{Dynamic Spin Structure Factor}

\begin{figure}[H]
    \centering
    \includegraphics[width=\columnwidth]{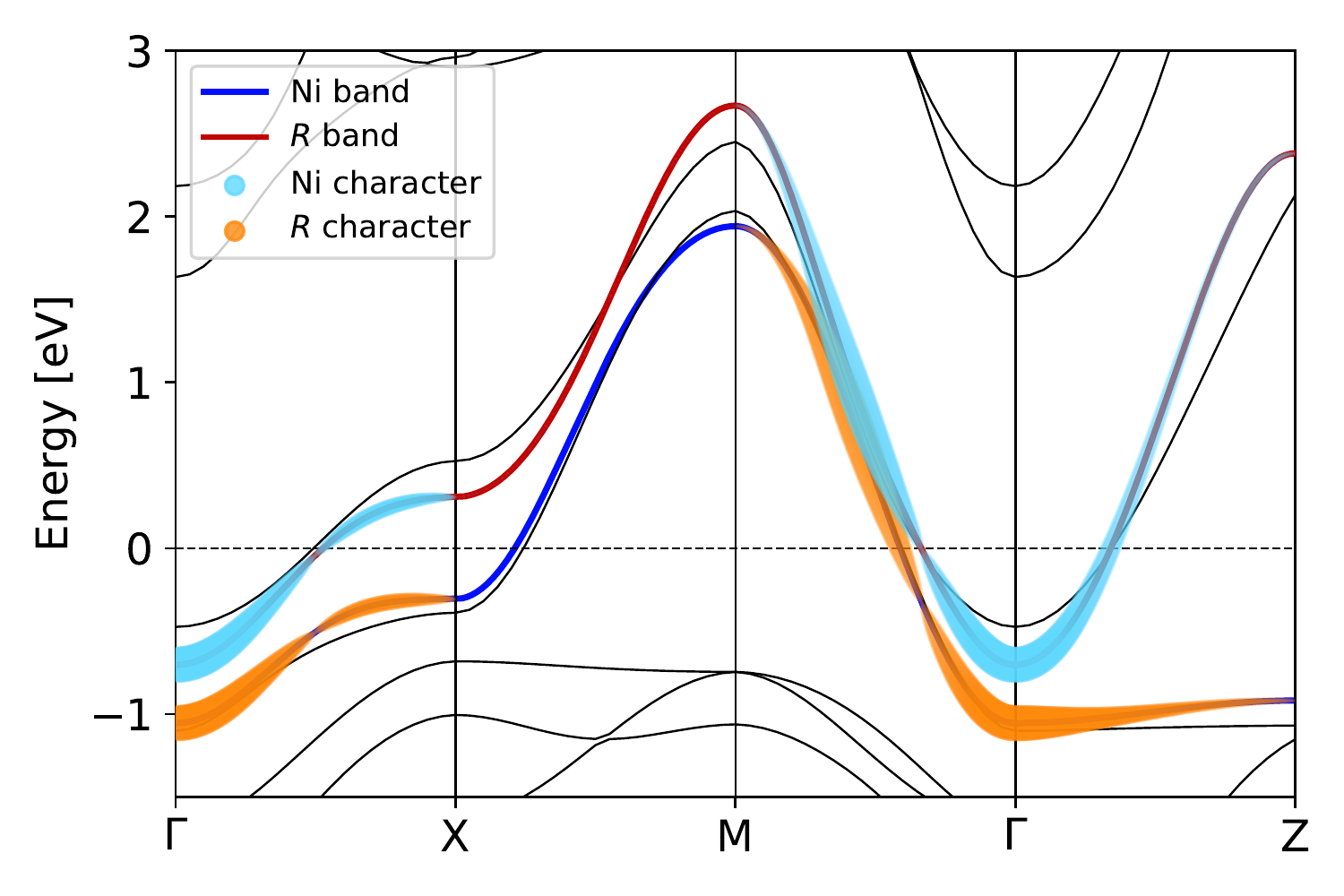}
    \caption{Hybridization of the two-orbital model for NdNiO$_2$, as described in Eq.~\ref{eq:Hubbard_Hamiltonian}, showing subdominant Ni and $R$ band character, respectively.}
    \label{fig:bandcharacter}
\end{figure}

The dynamic spin structure factor has been shown to be an accurate approximation of the RIXS cross section of both Mott insulators and doped systems \citep{Jia2014,Jia_2016}. The dynamical spin structure factor is defined as
\begin{equation}
\label{eq:Sqw}
    S(\textbf{q},\omega) = \frac{1}{\pi} {\rm Im} \left\langle G \left| \rho_{\textbf{q}}^{(s)\dagger} \frac{1}{\mathcal{H}-E_G - \omega - i\Gamma} \rho_{\textbf{q}}^{(s)} \right|G \right\rangle,
\end{equation}
where the spin density operator is given by $\rho_{\textbf{q}}^{(s)} = \sum_{\textbf{i},\sigma} s_{\textbf{i}\sigma} e^{i\textbf{q}\cdot\textbf{r}_{\textbf{i}}} = \sum_{\textbf{k},\sigma} \sigma c_{\textbf{k}+\textbf{q},\sigma}^{\dagger} c_{\textbf{k}\sigma}$,
%\begin{equation}
%    \rho_{\textbf{q}}^{(s)} = \sum_{\textbf{i},\sigma} s_{\textbf{i}\sigma} e^{i\textbf{q}\cdot\textbf{r}_{\textbf{i}}} = \sum_{\textbf{k},\sigma} \sigma c_{\textbf{k}+\textbf{q},\sigma}^{\dagger} c_{\textbf{k}\sigma}
%\end{equation}
and the two-band nickel model Hamiltonian is given by
\begin{equation}
\label{eq:Hubbard_Hamiltonian}
\begin{split}
    \mathcal{H} = \sum_{\textbf{k},\sigma} & \left( \varepsilon_{\textbf{k}}^R c^{\dagger}_{\textbf{k}\sigma}c_{\textbf{k}\sigma} + \varepsilon_{\textbf{k}}^{\rm Ni} d^{\dagger}_{\textbf{k}\sigma}d_{\textbf{k}\sigma} \right) \\
    &+ \frac{U}{N} \sum_{\textbf{k}_1, \textbf{k}_2, \textbf{q}} d_{\textbf{k}_1+\textbf{q},\uparrow}^{\dagger} d_{\textbf{k}_2-\textbf{q},\downarrow}^{\dagger} d_{\textbf{k}_2,\downarrow} d_{\textbf{k}_1,\uparrow} \\
    &+ \sum_{\textbf{k},\sigma}\left(\varepsilon_{\textbf{k}}^{R-Ni} c_{\textbf{k},\sigma}^{\dagger} d_{\textbf{k},\sigma} + {\rm H.c.} \right),
\end{split}
\end{equation}
where $c_{\textbf{k}}$ ($c^{\dagger}_{\textbf{k}}$) operators represent the dispersive $R$ $5d$ band and the $d_{\textbf{k}}$ ($d^{\dagger}_{\textbf{k}}$) operators represent the Hubbard-like Ni $3d$ band. The details of $\varepsilon^{R}_{\textbf{k}}$, $\varepsilon^{\rm Ni}_{\textbf{k}}$, and $V_{\textbf{k},\textbf{q}}$ are defined in \cite{Been_2021} which proposed this form of the two orbital model. The hybridization between Ni and $R$ in this 2-orbital model is shown visually by plotting only the subdominant orbital character in Figure \ref{fig:bandcharacter}. This hybridization results in the "self-doped" nature of the nickelates.

We evaluate the dynamic spin structure factor, $S(\textbf{q},\omega)$, for the two-band nickelate model using exact diagonalization (ED) on an 8-site (diamond-shaped) Betts cluster \citep{Betts_1999} with periodic boundary conditions, which suffers from finite-size effects but is sufficient for determining the magnon spectrum along the antiferromagnetic Brillouin zone at $(\pi/2,\pi/2)$. We take the value of the Hubbard $U=8$ eV so that we get a reasonable estimate of $J \sim 80$ meV. The eigenvalues and eigenvectors of the ED calculations were found using the Implicitly Restarted Lanczos Method from the ARnoldi PACKage (ARPACK), as implemented in SciPy Linalg library (\cite{2020SciPy-NMeth}, RRID: SCR\_008058). The biconjugate gradient stabilized method was used to calculate $S(\textbf{q},\omega)$.

\section{Results}
\subsection{XAS}
\subsubsection{Single-site}\label{singlesite}

The single site XAS is simulated using techniques mentioned in Sec.~\ref{sec:XAS}, where the Hilbert space is determined using only the 3d (valence) and 2p (core) orbitals of the Ni transition metal ion. The eigenenergies are obtained by diagonalizing the Hamiltonian and the XAS cross section is evaluated using Eq.~\ref{Eq:XAS}.

\begin{figure}[H]
    \begin{center}
    \includegraphics[width=\columnwidth,trim={60 0 65 0},clip]{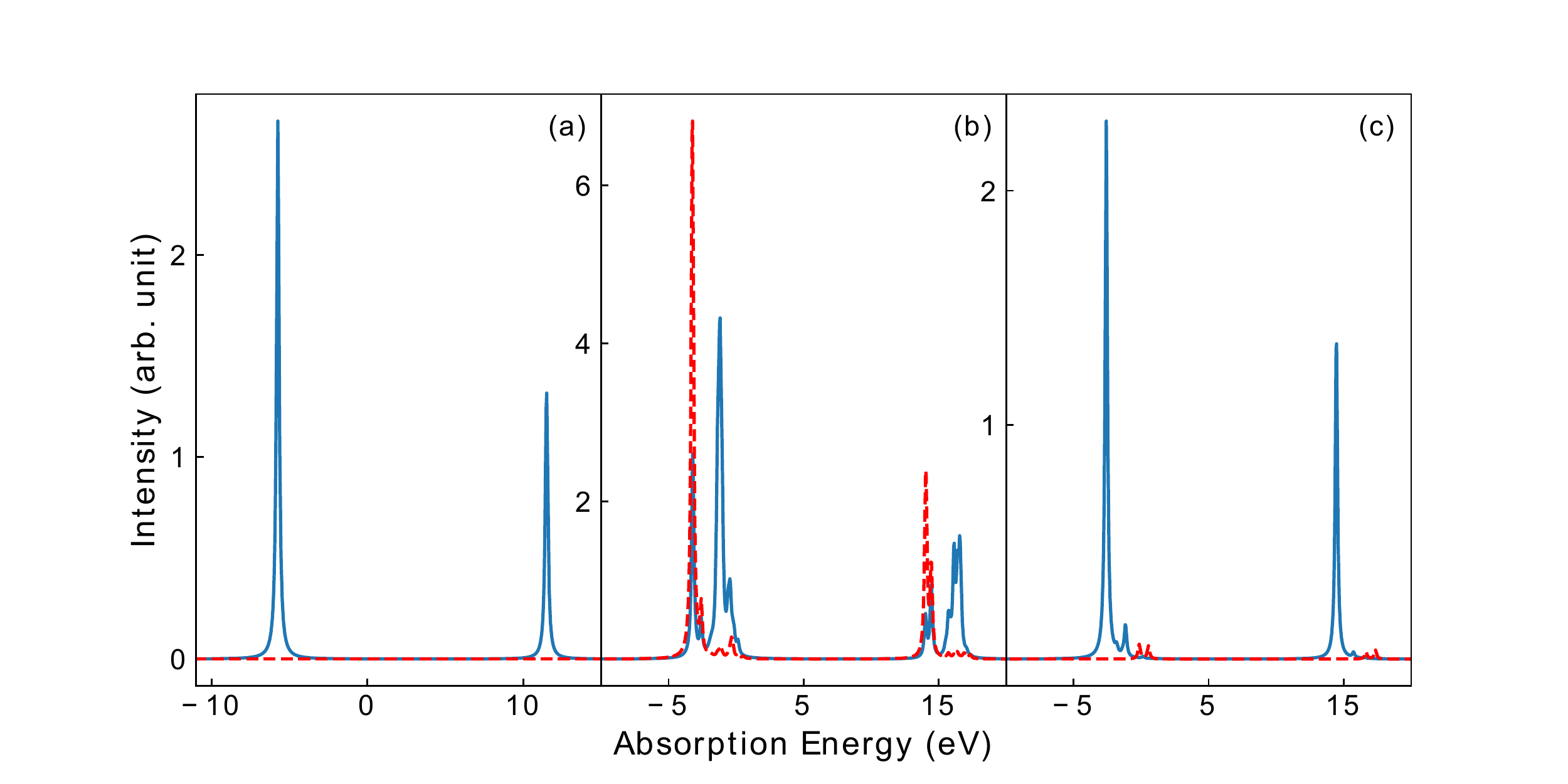}
    \end{center}
\caption{Multiplet XAS calculations for the single-site nickel ion. The absorption energy is measured with respect to the ground state of $d^9$ ion. The solid line is the spectra measured with $x$-polarized light, and the dashed line is the spectra measured with $z$-polarized light. From left to right are spectra of (a) the $d^9$ ion, (b) the $d^8$ high-spin ion, and (c) the $d^8$ low-spin ion.}\label{fig:1siteud}
\end{figure}

We first calculate the multiplet XAS of the $d^9$ electronic configuration of the Ni ion using the parameters from \cite{rossi2020orbital}, where the results are plotted in Fig.~\ref{fig:1siteud}. Since the single site calculation does not include ligand orbitals in the Hilbert space, direct measurements of crystal field splitting of $d$ orbitals energy levels are used without hybridization parameters obtained by Wannier downfolding from \cite{hepting2020electronic}. The Coulomb interaction for $d^9$ ions is less relevant and the spectral lineshape is dominated by the spin orbit coupling of the core levels, showing a single peak for both the $L_3$ and $L_2$ edges with light polarization along the x-direction, and no absorption with light polarization along the $z$-direction, due to the orbital $d_{x^2-y^2}$ character of the holes in the $d^9$ configuration. We then calculated the multiplet XAS on the $d^8$ ion using the same parameters for a spin singlet (S = 0) and a spin triplet (S = 1) ground state. Similar to previous reports \cite{rossi2020orbital}, with light polarized along the $z$-direction the intensity of the XAS peaks decrease in the low spin state, while the intensity remains the same order of magnitude for the high spin state. With light polarized along the $x$-direction, the high-spin ground state displays stronger intensity across a wide range of absorption energies, with more dipole-allowed transitions to excited states, compared to the singly degenerate, low-spin ground state.

\begin{figure}[H]
    \begin{center}
    \includegraphics[width=\columnwidth,trim={60 0 65 0},clip]{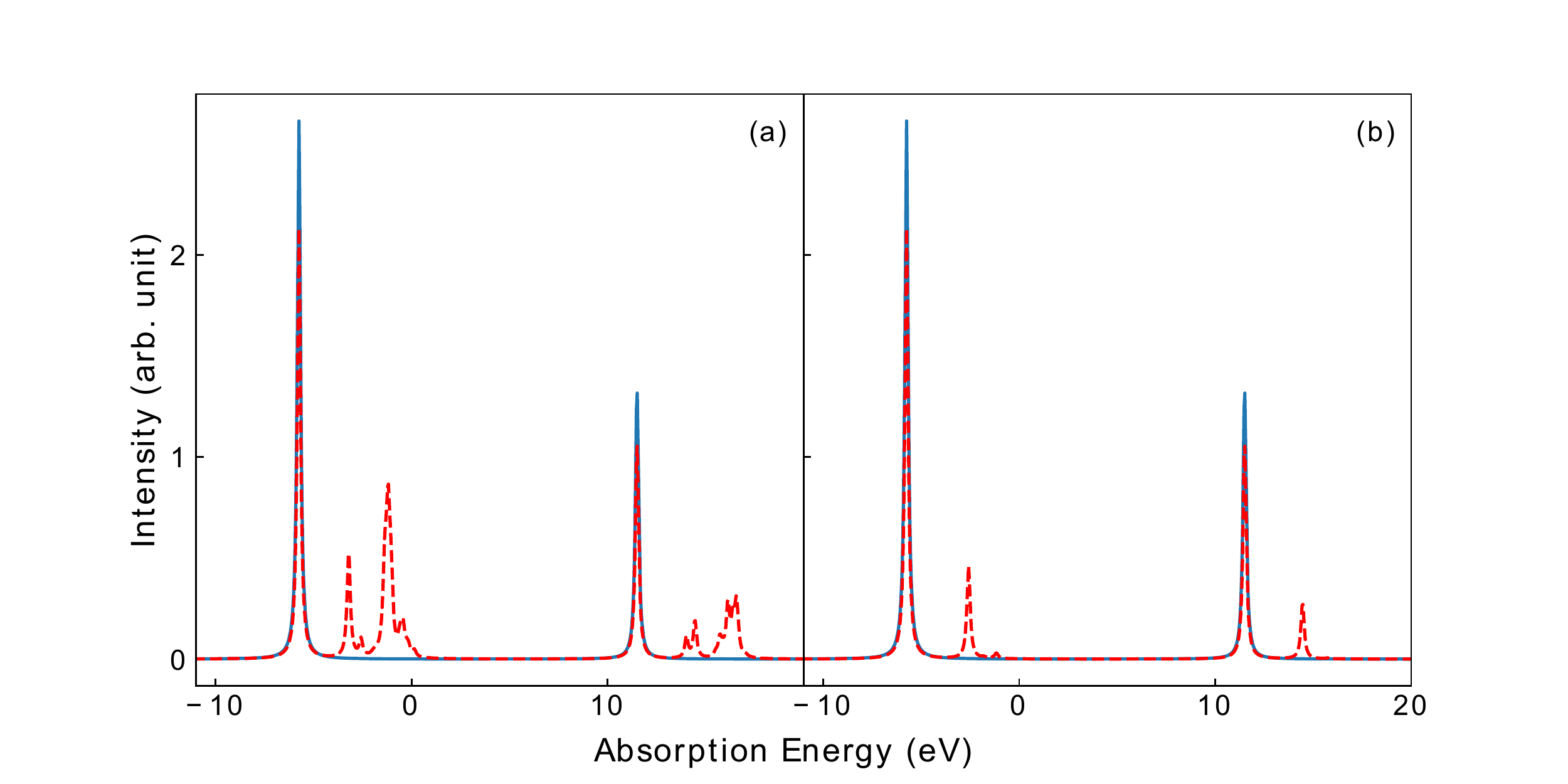}
    \end{center}
    \caption{XAS spectra for the undoped $d^9$ ion (solid line) and with 20\% of the hole-doped $d^8$ ion (dashed line) calculated for $x$-polarization. Panel (a) shows the results for doping with the $d^8$ high-spin state, and panel (b) shows the results for doping with the $d^8$ low-spin state.}
    \label{fig:1sitedoped}
\end{figure}

We also simulated doped spectra as a linear combination of the $d^8$ and $d^9$ spectra, with the results plotted in Fig.~\ref{fig:1sitedoped}. Clearly, the high-spin $d^8$ state produces spectra with a wide energy spread and additional multiplet peaks. The better comparison to experiment \citep{rossi2020orbital} can be made if the $d^8$ ion is in the low-spin configuration, producing a less pronounced shoulder in addition to the main absorption peak from the dominant $d^9$ configuration.

\subsubsection{Two-site}
The two-site XAS is simulated with a two-site, multi-orbital Hubbard model as described in Sec.~\ref{sec:XAS}, with 11 orbitals for each site (unit cell NiO$_2$) in the initial state that includes 6 oxygen $2p$ orbitals and the 5 nickel $3d$ orbitals , and 14 orbitals for one of the cells in the final state, which includes an additional 3 core-level nickel $2p$ orbitals for the target atom. The eigenenergies and eigenstates are obtained by exact diagonalization of the model Hamiltonian and the XAS spectra are calculated using Eq.~\ref{Eq:XAS}. Including the oxygen ligand orbitals in the model allows hybridization and charge transfer effects between oxygen $2p$ and Nickel $3d$ orbitals. The Ni-O hybridization will affect not only the effective energy levels of the Nickel $3d$ orbitals, but this also changes the effective band character, with oxygen ligand $p$ orbital contributions to the valance and conduction bands, which will result in changes to the XAS lineshape compared to the single-site calculation in Sec.~\ref{singlesite}.

\begin{figure}[h!]
\begin{center}
\includegraphics[width=\columnwidth,trim={60 0 80 0},clip]{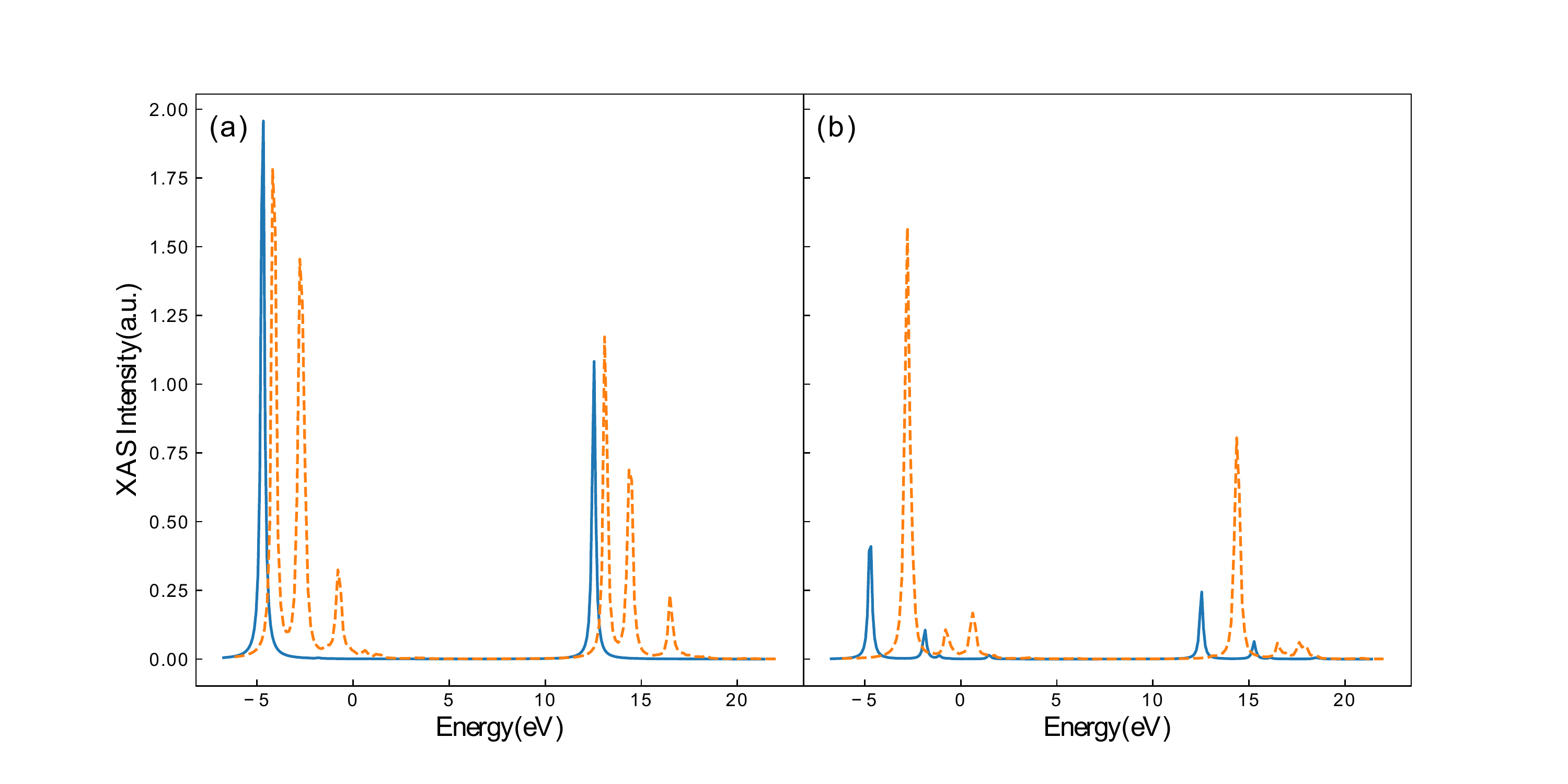}
\end{center}
\caption{Multiplet calculations for two-site, two-hole (solid, blue line) and two-site, three-hole (dashed, orange) NiO$_2$ clusters. The absorption energy is measured from the ground state of the two-site, two holes calculation. Panel (a) shows the XAS spectra in $x$-polarization; and panel (b) shows the XAS spectra in $z$-polarization.}\label{fig:2sites}
\end{figure}

We performed multiplet calculations for a two-site, two-hole (nominally $d^9$ nickel) and a two-site, three-hole (doped NiO$_2$ planes) cluster using effective parameters from \cite{rossi2020orbital}, and hybridization values obtained by Wannier downfolding as described in Sec.~\ref{sec:DFT}. The inclusion of hybridization parameters allow us to modulate the strength of electron hopping between each orbital and the onsite energy, with the electrons distributed between the different orbitals based on the model Hamiltonian as described in \cite{hepting2020electronic}. These XAS spectra are shown in Fig.~\ref{fig:2sites}. The two-site, two-hole XAS spectrum with $x$-polarization is similar to the single-site, $d^9$ model, which was described in the previous section. When doped with one hole, the $L_3$ and $L_2$ edges split into 3 peaks due to Ni-O hybridization. These results are consistent with both the single-site calculation and the experimental results \citep{rossi2020orbital}.%With $z$-polarization, the two-site, two-hole XAS spectrum has 3 peaks at each edge as shown in Fig.~\ref{fig:2sites}a. In contrast, the single-site, $d^9$ model with $z$ polarization doesn't have any peaks. To investigate the origin of the difference, we calculated the two-site, two-hole XAS spectrum without hybridization. The result shown in Fig. \ref{fig:2holes2sites} shows that there are no peaks in this case, which coincides the single-site model, indicating that the Ni-O hybridization leads to those peaks.

%We then calculated $S_z$ and hole occupations for the undoped and hole doped $NiO_2$ cluster to observe XAS spectra change if the Nickelate is doped with one hole for every two sites and investigate the effect of the hybridization between 2p oxygen orbitals and 3d Ni orbitals. {\bf for what reason?} For the undoped cluster, calculations show that $S_z$ is 0 for the ground state. 1.52 holes occupy the $d_{x^2-y^2}$ orbital of Ni and 0.47 holes occupy the $p_x$ orbital of O on x axis. For the one hole doped case, calculations show that S$_z$ is $\pm$0.5 for the ground state. 1.96 holes occupy the $d_{x^2-y^2}$ orbital of Ni and 1.02 holes occupy the $p_x$ orbital of O on the x axis. The ground state calculation showed that $d^8$ Nickelate has a spin-singlet (S = 0) ground state as confirmed in other calculations in the previous section and other reports. {\bf how does this relate to your XAS spectra? Also, can you work to make these sentences a bit less awkward and more direct?} 

\
\subsection{Spin structure factor}

\begin{figure*}
    \centering
    \includegraphics[width=\textwidth,trim={0 300 0 0},clip]{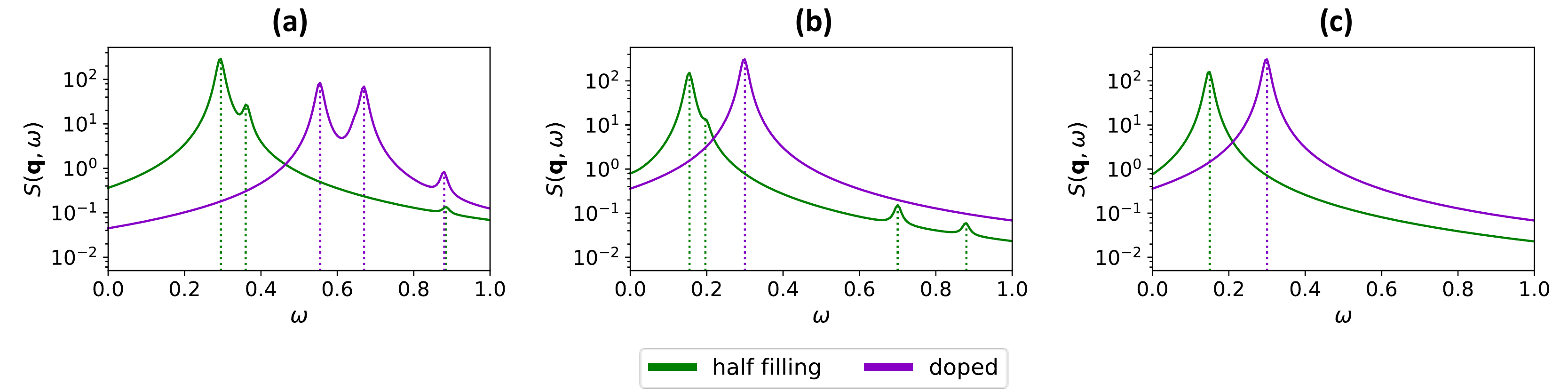}
    \caption{Comparison of $S(\textbf{q},\omega)$ at $\textbf{q}=\left(\frac{\pi}{2}, \frac{\pi}{2} \right)$ between \textbf{(a)} the two-orbital model with $\varepsilon_{\textbf{k}}^{R-Ni} \sim 1$ eV, \textbf{(b)} the two-orbital model with $\varepsilon_{\textbf{k}}^{R-Ni} \rightarrow \varepsilon_{\textbf{k}}^{R-Ni}+U/2$, and \textbf{(c)} the one-orbital model that only includes the Ni layer. The dotted vertical lines in each plot show positions of energy eigenvalues (differences) %that were fitted to the $S(\textbf{q},\omega)$ curves when 
    broadened by Lorentzian convolution in the final spectrum. All energies are in eV with $S(\textbf{q},\omega)$ shown on a log scale to enhance weaker features.}
    \label{fig:Sqw}
\end{figure*}

Results for the dynamical spin structure factor calculated via exact diagonalization are plotted in Fig.~\ref{fig:Sqw}, showing a comparison of the two-orbital model with two different values of the energy difference between Ni and $R$ orbitals $\varepsilon_{\textbf{k}}^{R-Ni}$, and also results from a single-orbital model for the nickel oxide layer. ``Half-filling" corresponds to 8 electrons distributed among the 16 orbitals of the two-orbital, 8A Betts cluster or the 8 orbitals of the 8A Betts cluster in the single-orbital model for the Ni-O layer. The Ni-$R$ energy difference and hybridization dictate the effective filling of the Ni band in the two-orbital model. The doped calculation is performed with 6 electrons, corresponding to a nominal doping of  $\sim 25$\%.

For the two-orbital model, tuning the energy difference between Ni and $R$ sites, $\varepsilon_{\textbf{k}}^{R-Ni}$, plays a major role in the occupation of the $R$ orbital, and hence the "self-doping" of the Ni layer. In Fig.~\ref{fig:Sqw}$(a)$, $\varepsilon_{\textbf{k}}^{R-Ni} \sim 1$ %which matches exactly the parameters predicted by Wannier downfolding in \cite{Been_2021}. When taken together with the total Ni-$R$ occupations in Table \ref{tab:Ni-R_occupation}, This small of a $\Delta$ 
results in approximately two electrons occupying the $R$-sites at both overall half-filling and $\sim 25$\% doping, meaning the Ni layer retains a 25\% ``self-doping" offset. In contrast, panel Fig.~\ref{fig:Sqw}\textbf{(b)} shows the results for the two-orbital model with a large offset of $U/2=4$eV added to $\varepsilon_{\textbf{k}}^{R-Ni}$, resulting in $\varepsilon_{\textbf{k}}^{R-Ni} \sim 5$ eV. Here, %referencing the occupation of the Ni and $R$ sites in Table \ref{tab:Ni-R_occupation}, 
the $R$-site is almost completely empty and $S(\textbf{q},\omega)$ looks very similar to results from the single-orbital calculation shown in Fig.~\ref{fig:Sqw}\textbf{(c)}. %, the major difference being some small satellites.
Obviously, the infinite-layer nickelates live between these two extremes and Fig.~\ref{fig:Sqw} gives us some clues as to how $S(\textbf{q},\omega)$ is affected by the occupation of the $R$ sites, with $S(\textbf{q},\omega)$ extremely sensitive to the $R$ and Ni occupations.

\section{Discussion}
In this paper we explored numerical simulations using DFT and ED to determine the underlying band structure, valence configuration, and dynamical spin response of model Hamiltonians aimed to describe infinite-layer nickelates. DFT for various $R$ substituted nickelates was used to determine the parameters of an effective two orbital model including Ni 3$d_{x^2-y^2}$ and $R$ 5$d$ "axial" orbitals, where the effect of lower-lying oxygen orbitals is to modify the inter-orbital hybridizations. Cluster multiplet ED is used to determine that undoped nickelate has predominantly 3$d^9$ valence which upon doping involves low-spin nickel 3$d^8$ valence holes. These results indicate that undoped infinite-layer nickelate is "self-doped" away from 1/2 filling Ni 3$d^9$ via the presence of a finite electron concentration in the $R$ layer compensating the holes in the Ni layer, the physics of doped nickelates may be described simply from the point of view of a one-band Hubbard-like model. Our consideration of the spin response indicates a close similarity of the paramagnon energies and intensities to one-band systems, in close analogy with the behavior seen in the cuprates.

Our ED calculations are limited to small clusters and therefore the effect of long-wavelength physics remains beyond our level of investigation. This may be particularly relevant to a discussion of the root cause of superconductivity and the competition between suoerconductivity and other intertwined phases, such as spin and/or charge stripes that are prevalent in both cuprate phase diagram and numerical simulations of the single-band Hubbard model \citep{Arovas,JPSJ}. \cite{Peng_DMRG_arxiv} reports density matrix renormalization group (DMRG) simulations of a similar two-orbital model, and finds Luther-Emery behavior - coexistence of long-range superconducting and charge-density wave order - away from half filling as in the 1D Hubbard model, while the undoped model does not contain long-range antiferromagnetic order, different than single-band Hubbard and more closely in-line with infinite-layer nickelates. The spin dynamics may be investigated using t-DMRG or other techniques, such as determinant quantum Monte Carlo \citep{Fangze}. This remains a topic of future interest.

\section*{Author Contributions}

E.M.B. performed ED calculations for $S(q,\omega)$, Y. H. and K. H. H. performed ED calculations for XAS, C.J. and E.M.B. performed DFT calculations for the nickelate bandstructure. B.M., C.J., Y.C., and T.P.D. conceived the project. All authors contributed to the writing of the manuscript.

\section*{Funding}
This work was supported by the U.S. Department of Energy (DOE), Office of Basic Energy Sciences,
Division of Materials Sciences and Engineering.
Computational work was performed on the Sherlock cluster at Stanford University and on resources of the National Energy Research Scientific Computing Center, supported by the U.S. DOE, Office of Science, under Contract no. DE-AC02-05CH11231.

\section*{Acknowledgments}
We would like to thank Professor Yao Wang from Clemson University for helpful discussions about exact diagonalization and the two-orbital nickelate model.

%\section*{Supplemental Data}
% \href{http://home.frontiersin.org/about/author-guidelines#SupplementaryMaterial}{Supplementary Material} should be uploaded separately on submission, if there are Supplementary Figures, please include the caption in the same file as the figure. LaTeX Supplementary Material templates can be found in the Frontiers LaTeX folder.

%\section*{Data Availability Statement}
%The datasets [GENERATED/ANALYZED] for this study can be found in the [NAME OF REPOSITORY] [LINK].
% Please see the availability of data guidelines for more information, at https://www.frontiersin.org/about/author-guidelines#AvailabilityofData

%\bibliographystyle{frontiersinSCNS_ENG_HUMS} % for Science, Engineering and Humanities and Social Sciences articles, for Humanities and Social Sciences articles please include page numbers in the in-text citations
%\bibliographystyle{frontiersinHLTH&FPHY} % for Health, Physics and Mathematics articles
\bibliography{biblio}

%merlin.mbs apsrev4-1.bst 2010-07-25 4.21a (PWD, AO, DPC) hacked
%Control: key (0)
%Control: author (0) dotless jnrlst
%Control: editor formatted (1) identically to author
%Control: production of article title (0) allowed
%Control: page (1) range
%Control: year (0) verbatim
%Control: production of eprint (0) enabled
\begin{thebibliography}{22}%
\makeatletter
\providecommand \@ifxundefined [1]{%
 \@ifx{#1\undefined}
}%
\providecommand \@ifnum [1]{%
 \ifnum #1\expandafter \@firstoftwo
 \else \expandafter \@secondoftwo
 \fi
}%
\providecommand \@ifx [1]{%
 \ifx #1\expandafter \@firstoftwo
 \else \expandafter \@secondoftwo
 \fi
}%
\providecommand \natexlab [1]{#1}%
\providecommand \enquote  [1]{``#1''}%
\providecommand \bibnamefont  [1]{#1}%
\providecommand \bibfnamefont [1]{#1}%
\providecommand \citenamefont [1]{#1}%
\providecommand \href@noop [0]{\@secondoftwo}%
\providecommand \href [0]{\begingroup \@sanitize@url \@href}%
\providecommand \@href[1]{\@@startlink{#1}\@@href}%
\providecommand \@@href[1]{\endgroup#1\@@endlink}%
\providecommand \@sanitize@url [0]{\catcode `\\12\catcode `\$12\catcode
  `\&12\catcode `\#12\catcode `\^12\catcode `\_12\catcode `\%12\relax}%
\providecommand \@@startlink[1]{}%
\providecommand \@@endlink[0]{}%
\providecommand \url  [0]{\begingroup\@sanitize@url \@url }%
\providecommand \@url [1]{\endgroup\@href {#1}{\urlprefix }}%
\providecommand \urlprefix  [0]{URL }%
\providecommand \Eprint [0]{\href }%
\providecommand \doibase [0]{http://dx.doi.org/}%
\providecommand \selectlanguage [0]{\@gobble}%
\providecommand \bibinfo  [0]{\@secondoftwo}%
\providecommand \bibfield  [0]{\@secondoftwo}%
\providecommand \translation [1]{[#1]}%
\providecommand \BibitemOpen [0]{}%
\providecommand \bibitemStop [0]{}%
\providecommand \bibitemNoStop [0]{.\EOS\space}%
\providecommand \EOS [0]{\spacefactor3000\relax}%
\providecommand \BibitemShut  [1]{\csname bibitem#1\endcsname}%
\let\auto@bib@innerbib\@empty
%</preamble>
\bibitem [{\citenamefont {Li}\ \emph {et~al.}(2019)\citenamefont {Li},
  \citenamefont {Lee}, \citenamefont {Wang}, \citenamefont {Osada},
  \citenamefont {Crossley}, \citenamefont {Lee}, \citenamefont {Cui},
  \citenamefont {Hikita},\ and\ \citenamefont
  {Hwang}}]{Denver_nickelateDiscovery_2019}%
  \BibitemOpen
  \bibfield  {author} {\bibinfo {author} {\bibfnamefont {Danfeng}\ \bibnamefont
  {Li}}, \bibinfo {author} {\bibfnamefont {Kyuho}\ \bibnamefont {Lee}},
  \bibinfo {author} {\bibfnamefont {Bai~Yang}\ \bibnamefont {Wang}}, \bibinfo
  {author} {\bibfnamefont {Motoki}\ \bibnamefont {Osada}}, \bibinfo {author}
  {\bibfnamefont {Samuel}\ \bibnamefont {Crossley}}, \bibinfo {author}
  {\bibfnamefont {Hye~Ryoung}\ \bibnamefont {Lee}}, \bibinfo {author}
  {\bibfnamefont {Yi}~\bibnamefont {Cui}}, \bibinfo {author} {\bibfnamefont
  {Yasuyuki}\ \bibnamefont {Hikita}}, \ and\ \bibinfo {author} {\bibfnamefont
  {Harold~Y.}\ \bibnamefont {Hwang}},\ }\bibfield  {title} {\enquote {\bibinfo
  {title} {Superconductivity in an infinite-layer nickelate},}\ }\href
  {\doibase 10.1038/s41586-019-1496-5} {\bibfield  {journal} {\bibinfo
  {journal} {Nature}\ }\textbf {\bibinfo {volume} {572}},\ \bibinfo {pages}
  {624--627} (\bibinfo {year} {2019})}\BibitemShut {NoStop}%
\bibitem [{\citenamefont {Anisimov}\ \emph {et~al.}(1999)\citenamefont
  {Anisimov}, \citenamefont {Bukhvalov},\ and\ \citenamefont
  {Rice}}]{Anisimov_1999}%
  \BibitemOpen
  \bibfield  {author} {\bibinfo {author} {\bibfnamefont {V.~I.}\ \bibnamefont
  {Anisimov}}, \bibinfo {author} {\bibfnamefont {D.}~\bibnamefont {Bukhvalov}},
  \ and\ \bibinfo {author} {\bibfnamefont {T.~M.}\ \bibnamefont {Rice}},\
  }\bibfield  {title} {\enquote {\bibinfo {title} {Electronic structure of
  possible nickelate analogs to the cuprates},}\ }\href {\doibase
  10.1103/PhysRevB.59.7901} {\bibfield  {journal} {\bibinfo  {journal}
  {Physical Review B}\ }\textbf {\bibinfo {volume} {59}},\ \bibinfo {pages}
  {7901--7906} (\bibinfo {year} {1999})}\BibitemShut {NoStop}%
\bibitem [{\citenamefont {Lee}\ and\ \citenamefont
  {Pickett}(2004)}]{Pickett_2004}%
  \BibitemOpen
  \bibfield  {author} {\bibinfo {author} {\bibfnamefont {K.-W.}\ \bibnamefont
  {Lee}}\ and\ \bibinfo {author} {\bibfnamefont {W.~E.}\ \bibnamefont
  {Pickett}},\ }\bibfield  {title} {\enquote {\bibinfo {title} {{Infinite-layer
  $\mathrm{La}\mathrm{Ni}{\mathrm{O}}_{2}$: ${\mathrm{Ni}}^{1+}$ is not
  ${\mathrm{Cu}}^{2+}$}},}\ }\href {\doibase 10.1103/PhysRevB.70.165109}
  {\bibfield  {journal} {\bibinfo  {journal} {Physical Review B}\ }\textbf
  {\bibinfo {volume} {70}},\ \bibinfo {pages} {165109} (\bibinfo {year}
  {2004})}\BibitemShut {NoStop}%
\bibitem [{\citenamefont {Zaanen}\ \emph {et~al.}(1985)\citenamefont {Zaanen},
  \citenamefont {Sawatzky},\ and\ \citenamefont {Allen}}]{ZSA}%
  \BibitemOpen
  \bibfield  {author} {\bibinfo {author} {\bibfnamefont {J.}~\bibnamefont
  {Zaanen}}, \bibinfo {author} {\bibfnamefont {G.~A.}\ \bibnamefont
  {Sawatzky}}, \ and\ \bibinfo {author} {\bibfnamefont {J.~W.}\ \bibnamefont
  {Allen}},\ }\bibfield  {title} {\enquote {\bibinfo {title} {{Band gaps and
  electronic structure of transition-metal compounds}},}\ }\href {\doibase
  10.1103/PhysRevLett.55.418} {\bibfield  {journal} {\bibinfo  {journal}
  {Physical Review Letters}\ }\textbf {\bibinfo {volume} {55}},\ \bibinfo
  {pages} {418--421} (\bibinfo {year} {1985})}\BibitemShut {NoStop}%
\bibitem [{\citenamefont {Lu}\ \emph {et~al.}(2021)\citenamefont {Lu},
  \citenamefont {Rossi}, \citenamefont {Nag}, \citenamefont {Osada},
  \citenamefont {Li}, \citenamefont {Lee}, \citenamefont {Wang}, \citenamefont
  {Garcia-Fernandez}, \citenamefont {Agrestini}, \citenamefont {Shen},
  \citenamefont {Been}, \citenamefont {Moritz}, \citenamefont {Devereaux},
  \citenamefont {Zaanen}, \citenamefont {Hwang}, \citenamefont {Zhou},\ and\
  \citenamefont {Lee}}]{Lu_2021}%
  \BibitemOpen
  \bibfield  {author} {\bibinfo {author} {\bibfnamefont {H.}~\bibnamefont
  {Lu}}, \bibinfo {author} {\bibfnamefont {M.}~\bibnamefont {Rossi}}, \bibinfo
  {author} {\bibfnamefont {A.}~\bibnamefont {Nag}}, \bibinfo {author}
  {\bibfnamefont {M.}~\bibnamefont {Osada}}, \bibinfo {author} {\bibfnamefont
  {D.~F.}\ \bibnamefont {Li}}, \bibinfo {author} {\bibfnamefont
  {K.}~\bibnamefont {Lee}}, \bibinfo {author} {\bibfnamefont {B.~Y.}\
  \bibnamefont {Wang}}, \bibinfo {author} {\bibfnamefont {M.}~\bibnamefont
  {Garcia-Fernandez}}, \bibinfo {author} {\bibfnamefont {S.}~\bibnamefont
  {Agrestini}}, \bibinfo {author} {\bibfnamefont {Z.~X.}\ \bibnamefont {Shen}},
  \bibinfo {author} {\bibfnamefont {E.~M.}\ \bibnamefont {Been}}, \bibinfo
  {author} {\bibfnamefont {B.}~\bibnamefont {Moritz}}, \bibinfo {author}
  {\bibfnamefont {T.~P.}\ \bibnamefont {Devereaux}}, \bibinfo {author}
  {\bibfnamefont {J.}~\bibnamefont {Zaanen}}, \bibinfo {author} {\bibfnamefont
  {H.~Y.}\ \bibnamefont {Hwang}}, \bibinfo {author} {\bibfnamefont {Ke-Jin}\
  \bibnamefont {Zhou}}, \ and\ \bibinfo {author} {\bibfnamefont {W.~S.}\
  \bibnamefont {Lee}},\ }\bibfield  {title} {\enquote {\bibinfo {title}
  {Magnetic excitations in infinite-layer nickelates},}\ }\href {\doibase
  10.1126/science.abd7726} {\bibfield  {journal} {\bibinfo  {journal}
  {Science}\ }\textbf {\bibinfo {volume} {373}},\ \bibinfo {pages} {213--216}
  (\bibinfo {year} {2021})},\ \Eprint
  {http://arxiv.org/abs/https://science.sciencemag.org/content/373/6551/213.full.pdf}
  {https://science.sciencemag.org/content/373/6551/213.full.pdf} \BibitemShut
  {NoStop}%
\bibitem [{\citenamefont {Jia}\ \emph {et~al.}(2014)\citenamefont {Jia},
  \citenamefont {Nowadnick}, \citenamefont {Wohlfeld}, \citenamefont {Kung},
  \citenamefont {Chen}, \citenamefont {Johnston}, \citenamefont {Tohyama},
  \citenamefont {Moritz},\ and\ \citenamefont {Devereaux}}]{Jia2014}%
  \BibitemOpen
  \bibfield  {author} {\bibinfo {author} {\bibfnamefont {C.~J.}\ \bibnamefont
  {Jia}}, \bibinfo {author} {\bibfnamefont {E.~A.}\ \bibnamefont {Nowadnick}},
  \bibinfo {author} {\bibfnamefont {K.}~\bibnamefont {Wohlfeld}}, \bibinfo
  {author} {\bibfnamefont {Y.~F.}\ \bibnamefont {Kung}}, \bibinfo {author}
  {\bibfnamefont {C.-C.}\ \bibnamefont {Chen}}, \bibinfo {author}
  {\bibfnamefont {S.}~\bibnamefont {Johnston}}, \bibinfo {author}
  {\bibfnamefont {T.}~\bibnamefont {Tohyama}}, \bibinfo {author} {\bibfnamefont
  {B.}~\bibnamefont {Moritz}}, \ and\ \bibinfo {author} {\bibfnamefont {T.~P.}\
  \bibnamefont {Devereaux}},\ }\bibfield  {title} {\enquote {\bibinfo {title}
  {Persistent spin excitations in doped antiferromagnets revealed by resonant
  inelastic light scattering},}\ }\href {\doibase 10.1038/ncomms4314}
  {\bibfield  {journal} {\bibinfo  {journal} {Nature Communications}\ }\textbf
  {\bibinfo {volume} {5}},\ \bibinfo {pages} {3314} (\bibinfo {year}
  {2014})}\BibitemShut {NoStop}%
\bibitem [{\citenamefont {{Le Tacon}}\ \emph {et~al.}(2011)\citenamefont {{Le
  Tacon}}, \citenamefont {{Ghiringhelli}}, \citenamefont {{Chaloupka}},
  \citenamefont {{Sala}}, \citenamefont {{Hinkov}}, \citenamefont
  {{Haverkort}}, \citenamefont {{Minola}}, \citenamefont {{Bakr}},
  \citenamefont {{Zhou}}, \citenamefont {{Blanco-Canosa}}, \citenamefont
  {{Monney}}, \citenamefont {{Song}}, \citenamefont {{Sun}}, \citenamefont
  {{Lin}}, \citenamefont {{de Luca}}, \citenamefont {{Salluzzo}}, \citenamefont
  {{Khaliullin}}, \citenamefont {{Schmitt}}, \citenamefont {{Braicovich}},\
  and\ \citenamefont {{Keimer}}}]{LeTacon_2011}%
  \BibitemOpen
  \bibfield  {author} {\bibinfo {author} {\bibfnamefont {M.}~\bibnamefont {{Le
  Tacon}}}, \bibinfo {author} {\bibfnamefont {G.}~\bibnamefont
  {{Ghiringhelli}}}, \bibinfo {author} {\bibfnamefont {J.}~\bibnamefont
  {{Chaloupka}}}, \bibinfo {author} {\bibfnamefont {M.~Moretti}\ \bibnamefont
  {{Sala}}}, \bibinfo {author} {\bibfnamefont {V.}~\bibnamefont {{Hinkov}}},
  \bibinfo {author} {\bibfnamefont {M.~W.}\ \bibnamefont {{Haverkort}}},
  \bibinfo {author} {\bibfnamefont {M.}~\bibnamefont {{Minola}}}, \bibinfo
  {author} {\bibfnamefont {M.}~\bibnamefont {{Bakr}}}, \bibinfo {author}
  {\bibfnamefont {K.~J.}\ \bibnamefont {{Zhou}}}, \bibinfo {author}
  {\bibfnamefont {S.}~\bibnamefont {{Blanco-Canosa}}}, \bibinfo {author}
  {\bibfnamefont {C.}~\bibnamefont {{Monney}}}, \bibinfo {author}
  {\bibfnamefont {Y.~T.}\ \bibnamefont {{Song}}}, \bibinfo {author}
  {\bibfnamefont {G.~L.}\ \bibnamefont {{Sun}}}, \bibinfo {author}
  {\bibfnamefont {C.~T.}\ \bibnamefont {{Lin}}}, \bibinfo {author}
  {\bibfnamefont {G.~M.}\ \bibnamefont {{de Luca}}}, \bibinfo {author}
  {\bibfnamefont {M.}~\bibnamefont {{Salluzzo}}}, \bibinfo {author}
  {\bibfnamefont {G.}~\bibnamefont {{Khaliullin}}}, \bibinfo {author}
  {\bibfnamefont {T.}~\bibnamefont {{Schmitt}}}, \bibinfo {author}
  {\bibfnamefont {L.}~\bibnamefont {{Braicovich}}}, \ and\ \bibinfo {author}
  {\bibfnamefont {B.}~\bibnamefont {{Keimer}}},\ }\bibfield  {title} {\enquote
  {\bibinfo {title} {{Intense paramagnon excitations in a large family of
  high-temperature superconductors}},}\ }\href {\doibase 10.1038/nphys2041}
  {\bibfield  {journal} {\bibinfo  {journal} {Nature Physics}\ }\textbf
  {\bibinfo {volume} {7}},\ \bibinfo {pages} {725--730} (\bibinfo {year}
  {2011})},\ \Eprint {http://arxiv.org/abs/1106.2641} {arXiv:1106.2641
  [cond-mat.supr-con]} \BibitemShut {NoStop}%
\bibitem [{\citenamefont {{Dean}}\ \emph {et~al.}(2013)\citenamefont {{Dean}},
  \citenamefont {{Dellea}}, \citenamefont {{Springell}}, \citenamefont
  {{Yakhou-Harris}}, \citenamefont {{Kummer}}, \citenamefont {{Brookes}},
  \citenamefont {{Liu}}, \citenamefont {{Sun}}, \citenamefont {{Strle}},
  \citenamefont {{Schmitt}}, \citenamefont {{Braicovich}}, \citenamefont
  {{Ghiringhelli}}, \citenamefont {{Bo{\v{z}}ovi{\'c}}},\ and\ \citenamefont
  {{Hill}}}]{Dean_2013}%
  \BibitemOpen
  \bibfield  {author} {\bibinfo {author} {\bibfnamefont {M.~P.~M.}\
  \bibnamefont {{Dean}}}, \bibinfo {author} {\bibfnamefont {G.}~\bibnamefont
  {{Dellea}}}, \bibinfo {author} {\bibfnamefont {R.~S.}\ \bibnamefont
  {{Springell}}}, \bibinfo {author} {\bibfnamefont {F.}~\bibnamefont
  {{Yakhou-Harris}}}, \bibinfo {author} {\bibfnamefont {K.}~\bibnamefont
  {{Kummer}}}, \bibinfo {author} {\bibfnamefont {N.~B.}\ \bibnamefont
  {{Brookes}}}, \bibinfo {author} {\bibfnamefont {X.}~\bibnamefont {{Liu}}},
  \bibinfo {author} {\bibfnamefont {Y.~J.}\ \bibnamefont {{Sun}}}, \bibinfo
  {author} {\bibfnamefont {J.}~\bibnamefont {{Strle}}}, \bibinfo {author}
  {\bibfnamefont {T.}~\bibnamefont {{Schmitt}}}, \bibinfo {author}
  {\bibfnamefont {L.}~\bibnamefont {{Braicovich}}}, \bibinfo {author}
  {\bibfnamefont {G.}~\bibnamefont {{Ghiringhelli}}}, \bibinfo {author}
  {\bibfnamefont {I.}~\bibnamefont {{Bo{\v{z}}ovi{\'c}}}}, \ and\ \bibinfo
  {author} {\bibfnamefont {J.~P.}\ \bibnamefont {{Hill}}},\ }\bibfield  {title}
  {\enquote {\bibinfo {title} {{Persistence of magnetic excitations in
  La$_{2-x}$Sr$_{x}$CuO$_{4}$ from the undoped insulator to the heavily
  overdoped non-superconducting metal}},}\ }\href {\doibase 10.1038/nmat3723}
  {\bibfield  {journal} {\bibinfo  {journal} {Nature Materials}\ }\textbf
  {\bibinfo {volume} {12}},\ \bibinfo {pages} {1019--1023} (\bibinfo {year}
  {2013})},\ \Eprint {http://arxiv.org/abs/1303.5359} {arXiv:1303.5359
  [cond-mat.supr-con]} \BibitemShut {NoStop}%
\bibitem [{\citenamefont {Giannozzi}\ \emph {et~al.}(2020)\citenamefont
  {Giannozzi}, \citenamefont {Baseggio}, \citenamefont {Bonfà}, \citenamefont
  {Brunato}, \citenamefont {Car}, \citenamefont {Carnimeo}, \citenamefont
  {Cavazzoni}, \citenamefont {de~Gironcoli}, \citenamefont {Delugas},
  \citenamefont {Ferrari~Ruffino}, \citenamefont {Ferretti}, \citenamefont
  {Marzari}, \citenamefont {Timrov}, \citenamefont {Urru},\ and\ \citenamefont
  {Baroni}}]{QE}%
  \BibitemOpen
  \bibfield  {author} {\bibinfo {author} {\bibfnamefont {Paolo}\ \bibnamefont
  {Giannozzi}}, \bibinfo {author} {\bibfnamefont {Oscar}\ \bibnamefont
  {Baseggio}}, \bibinfo {author} {\bibfnamefont {Pietro}\ \bibnamefont
  {Bonfà}}, \bibinfo {author} {\bibfnamefont {Davide}\ \bibnamefont
  {Brunato}}, \bibinfo {author} {\bibfnamefont {Roberto}\ \bibnamefont {Car}},
  \bibinfo {author} {\bibfnamefont {Ivan}\ \bibnamefont {Carnimeo}}, \bibinfo
  {author} {\bibfnamefont {Carlo}\ \bibnamefont {Cavazzoni}}, \bibinfo {author}
  {\bibfnamefont {Stefano}\ \bibnamefont {de~Gironcoli}}, \bibinfo {author}
  {\bibfnamefont {Pietro}\ \bibnamefont {Delugas}}, \bibinfo {author}
  {\bibfnamefont {Fabrizio}\ \bibnamefont {Ferrari~Ruffino}}, \bibinfo {author}
  {\bibfnamefont {Andrea}\ \bibnamefont {Ferretti}}, \bibinfo {author}
  {\bibfnamefont {Nicola}\ \bibnamefont {Marzari}}, \bibinfo {author}
  {\bibfnamefont {Iurii}\ \bibnamefont {Timrov}}, \bibinfo {author}
  {\bibfnamefont {Andrea}\ \bibnamefont {Urru}}, \ and\ \bibinfo {author}
  {\bibfnamefont {Stefano}\ \bibnamefont {Baroni}},\ }\bibfield  {title}
  {\enquote {\bibinfo {title} {{Quantum ESPRESSO toward the exascale}},}\
  }\href {\doibase 10.1063/5.0005082} {\bibfield  {journal} {\bibinfo
  {journal} {The Journal of Chemical Physics}\ }\textbf {\bibinfo {volume}
  {152}},\ \bibinfo {pages} {154105} (\bibinfo {year} {2020})},\ \Eprint
  {http://arxiv.org/abs/https://doi.org/10.1063/5.0005082}
  {https://doi.org/10.1063/5.0005082} \BibitemShut {NoStop}%
\bibitem [{\citenamefont {Perdew}\ \emph {et~al.}(1996)\citenamefont {Perdew},
  \citenamefont {Burke},\ and\ \citenamefont {Ernzerhof}}]{PBE}%
  \BibitemOpen
  \bibfield  {author} {\bibinfo {author} {\bibfnamefont {John~P.}\ \bibnamefont
  {Perdew}}, \bibinfo {author} {\bibfnamefont {Kieron}\ \bibnamefont {Burke}},
  \ and\ \bibinfo {author} {\bibfnamefont {Matthias}\ \bibnamefont
  {Ernzerhof}},\ }\bibfield  {title} {\enquote {\bibinfo {title} {Generalized
  gradient approximation made simple},}\ }\href {\doibase
  10.1103/PhysRevLett.77.3865} {\bibfield  {journal} {\bibinfo  {journal}
  {Physical Review Letters}\ }\textbf {\bibinfo {volume} {77}},\ \bibinfo
  {pages} {3865--3868} (\bibinfo {year} {1996})}\BibitemShut {NoStop}%
\bibitem [{\citenamefont {Mostofi}\ \emph {et~al.}(2014)\citenamefont
  {Mostofi}, \citenamefont {Yates}, \citenamefont {Pizzi}, \citenamefont {Lee},
  \citenamefont {Souza}, \citenamefont {Vanderbilt},\ and\ \citenamefont
  {Marzari}}]{Wannier90}%
  \BibitemOpen
  \bibfield  {author} {\bibinfo {author} {\bibfnamefont {Arash~A.}\
  \bibnamefont {Mostofi}}, \bibinfo {author} {\bibfnamefont {Jonathan~R.}\
  \bibnamefont {Yates}}, \bibinfo {author} {\bibfnamefont {Giovanni}\
  \bibnamefont {Pizzi}}, \bibinfo {author} {\bibfnamefont {Young-Su}\
  \bibnamefont {Lee}}, \bibinfo {author} {\bibfnamefont {Ivo}\ \bibnamefont
  {Souza}}, \bibinfo {author} {\bibfnamefont {David}\ \bibnamefont
  {Vanderbilt}}, \ and\ \bibinfo {author} {\bibfnamefont {Nicola}\ \bibnamefont
  {Marzari}},\ }\bibfield  {title} {\enquote {\bibinfo {title} {{An updated
  version of Wannier90: A tool for obtaining maximally-localised Wannier
  functions}},}\ }\href {\doibase https://doi.org/10.1016/j.cpc.2014.05.003}
  {\bibfield  {journal} {\bibinfo  {journal} {Computer Physics Communications}\
  }\textbf {\bibinfo {volume} {185}},\ \bibinfo {pages} {2309--2310} (\bibinfo
  {year} {2014})}\BibitemShut {NoStop}%
\bibitem [{\citenamefont {de~Groot}\ and\ \citenamefont
  {Kotani}(2008)}]{deGroot_book}%
  \BibitemOpen
  \bibfield  {author} {\bibinfo {author} {\bibfnamefont {F.}~\bibnamefont
  {de~Groot}}\ and\ \bibinfo {author} {\bibfnamefont {A.}~\bibnamefont
  {Kotani}},\ }\href {https://books.google.com/books?id=HGHzu66i1yoC} {\emph
  {\bibinfo {title} {Core Level Spectroscopy of Solids}}},\ Advances in
  Condensed Matter Science\ (\bibinfo  {publisher} {CRC Press},\ \bibinfo
  {year} {2008})\BibitemShut {NoStop}%
\bibitem [{\citenamefont {Jia}\ \emph {et~al.}(2016)\citenamefont {Jia},
  \citenamefont {Wohlfeld}, \citenamefont {Wang}, \citenamefont {Moritz},\ and\
  \citenamefont {Devereaux}}]{Jia_2016}%
  \BibitemOpen
  \bibfield  {author} {\bibinfo {author} {\bibfnamefont {Chunjing}\
  \bibnamefont {Jia}}, \bibinfo {author} {\bibfnamefont {Krzysztof}\
  \bibnamefont {Wohlfeld}}, \bibinfo {author} {\bibfnamefont {Yao}\
  \bibnamefont {Wang}}, \bibinfo {author} {\bibfnamefont {Brian}\ \bibnamefont
  {Moritz}}, \ and\ \bibinfo {author} {\bibfnamefont {Thomas~P.}\ \bibnamefont
  {Devereaux}},\ }\bibfield  {title} {\enquote {\bibinfo {title} {{Using RIXS
  to Uncover Elementary Charge and Spin Excitations}},}\ }\href {\doibase
  10.1103/PhysRevX.6.021020} {\bibfield  {journal} {\bibinfo  {journal}
  {Physical Review X}\ }\textbf {\bibinfo {volume} {6}},\ \bibinfo {pages}
  {021020} (\bibinfo {year} {2016})}\BibitemShut {NoStop}%
\bibitem [{\citenamefont {Been}\ \emph {et~al.}(2021)\citenamefont {Been},
  \citenamefont {Lee}, \citenamefont {Hwang}, \citenamefont {Cui},
  \citenamefont {Zaanen}, \citenamefont {Devereaux}, \citenamefont {Moritz},\
  and\ \citenamefont {Jia}}]{Been_2021}%
  \BibitemOpen
  \bibfield  {author} {\bibinfo {author} {\bibfnamefont {Emily}\ \bibnamefont
  {Been}}, \bibinfo {author} {\bibfnamefont {Wei-Sheng}\ \bibnamefont {Lee}},
  \bibinfo {author} {\bibfnamefont {Harold~Y.}\ \bibnamefont {Hwang}}, \bibinfo
  {author} {\bibfnamefont {Yi}~\bibnamefont {Cui}}, \bibinfo {author}
  {\bibfnamefont {Jan}\ \bibnamefont {Zaanen}}, \bibinfo {author}
  {\bibfnamefont {Thomas}\ \bibnamefont {Devereaux}}, \bibinfo {author}
  {\bibfnamefont {Brian}\ \bibnamefont {Moritz}}, \ and\ \bibinfo {author}
  {\bibfnamefont {Chunjing}\ \bibnamefont {Jia}},\ }\bibfield  {title}
  {\enquote {\bibinfo {title} {Electronic structure trends across the
  rare-earth series in superconducting infinite-layer nickelates},}\ }\href
  {\doibase 10.1103/PhysRevX.11.011050} {\bibfield  {journal} {\bibinfo
  {journal} {Physical Review X}\ }\textbf {\bibinfo {volume} {11}},\ \bibinfo
  {pages} {011050} (\bibinfo {year} {2021})}\BibitemShut {NoStop}%
\bibitem [{\citenamefont {Betts}\ \emph {et~al.}(1999)\citenamefont {Betts},
  \citenamefont {Lin},\ and\ \citenamefont {Flynn}}]{Betts_1999}%
  \BibitemOpen
  \bibfield  {author} {\bibinfo {author} {\bibfnamefont {D~D}\ \bibnamefont
  {Betts}}, \bibinfo {author} {\bibfnamefont {H~Q}\ \bibnamefont {Lin}}, \ and\
  \bibinfo {author} {\bibfnamefont {J~S}\ \bibnamefont {Flynn}},\ }\bibfield
  {title} {\enquote {\bibinfo {title} {Improved finite-lattice estimates of the
  properties of two quantum spin models on the infinite square lattice},}\
  }\href {\doibase 10.1139/p99-041} {\bibfield  {journal} {\bibinfo  {journal}
  {Canadian Journal of Physics}\ }\textbf {\bibinfo {volume} {77}},\ \bibinfo
  {pages} {353--369} (\bibinfo {year} {1999})},\ \Eprint
  {http://arxiv.org/abs/https://doi.org/10.1139/p99-041}
  {https://doi.org/10.1139/p99-041} \BibitemShut {NoStop}%
\bibitem [{\citenamefont {Virtanen}\ \emph {et~al.}(2020)\citenamefont
  {Virtanen}, \citenamefont {Gommers}, \citenamefont {Oliphant}, \citenamefont
  {Haberland}, \citenamefont {Reddy}, \citenamefont {Cournapeau}, \citenamefont
  {Burovski}, \citenamefont {Peterson}, \citenamefont {Weckesser},
  \citenamefont {Bright}, \citenamefont {{van der Walt}}, \citenamefont
  {Brett}, \citenamefont {Wilson}, \citenamefont {Millman}, \citenamefont
  {Mayorov}, \citenamefont {Nelson}, \citenamefont {Jones}, \citenamefont
  {Kern}, \citenamefont {Larson}, \citenamefont {Carey}, \citenamefont {Polat},
  \citenamefont {Feng}, \citenamefont {Moore}, \citenamefont {{VanderPlas}},
  \citenamefont {Laxalde}, \citenamefont {Perktold}, \citenamefont {Cimrman},
  \citenamefont {Henriksen}, \citenamefont {Quintero}, \citenamefont {Harris},
  \citenamefont {Archibald}, \citenamefont {Ribeiro}, \citenamefont
  {Pedregosa}, \citenamefont {{van Mulbregt}},\ and\ \citenamefont {{SciPy 1.0
  Contributors}}}]{2020SciPy-NMeth}%
  \BibitemOpen
  \bibfield  {author} {\bibinfo {author} {\bibfnamefont {Pauli}\ \bibnamefont
  {Virtanen}}, \bibinfo {author} {\bibfnamefont {Ralf}\ \bibnamefont
  {Gommers}}, \bibinfo {author} {\bibfnamefont {Travis~E.}\ \bibnamefont
  {Oliphant}}, \bibinfo {author} {\bibfnamefont {Matt}\ \bibnamefont
  {Haberland}}, \bibinfo {author} {\bibfnamefont {Tyler}\ \bibnamefont
  {Reddy}}, \bibinfo {author} {\bibfnamefont {David}\ \bibnamefont
  {Cournapeau}}, \bibinfo {author} {\bibfnamefont {Evgeni}\ \bibnamefont
  {Burovski}}, \bibinfo {author} {\bibfnamefont {Pearu}\ \bibnamefont
  {Peterson}}, \bibinfo {author} {\bibfnamefont {Warren}\ \bibnamefont
  {Weckesser}}, \bibinfo {author} {\bibfnamefont {Jonathan}\ \bibnamefont
  {Bright}}, \bibinfo {author} {\bibfnamefont {St{\'e}fan~J.}\ \bibnamefont
  {{van der Walt}}}, \bibinfo {author} {\bibfnamefont {Matthew}\ \bibnamefont
  {Brett}}, \bibinfo {author} {\bibfnamefont {Joshua}\ \bibnamefont {Wilson}},
  \bibinfo {author} {\bibfnamefont {K.~Jarrod}\ \bibnamefont {Millman}},
  \bibinfo {author} {\bibfnamefont {Nikolay}\ \bibnamefont {Mayorov}}, \bibinfo
  {author} {\bibfnamefont {Andrew R.~J.}\ \bibnamefont {Nelson}}, \bibinfo
  {author} {\bibfnamefont {Eric}\ \bibnamefont {Jones}}, \bibinfo {author}
  {\bibfnamefont {Robert}\ \bibnamefont {Kern}}, \bibinfo {author}
  {\bibfnamefont {Eric}\ \bibnamefont {Larson}}, \bibinfo {author}
  {\bibfnamefont {C~J}\ \bibnamefont {Carey}}, \bibinfo {author} {\bibfnamefont
  {{\.I}lhan}\ \bibnamefont {Polat}}, \bibinfo {author} {\bibfnamefont
  {Yu}~\bibnamefont {Feng}}, \bibinfo {author} {\bibfnamefont {Eric~W.}\
  \bibnamefont {Moore}}, \bibinfo {author} {\bibfnamefont {Jake}\ \bibnamefont
  {{VanderPlas}}}, \bibinfo {author} {\bibfnamefont {Denis}\ \bibnamefont
  {Laxalde}}, \bibinfo {author} {\bibfnamefont {Josef}\ \bibnamefont
  {Perktold}}, \bibinfo {author} {\bibfnamefont {Robert}\ \bibnamefont
  {Cimrman}}, \bibinfo {author} {\bibfnamefont {Ian}\ \bibnamefont
  {Henriksen}}, \bibinfo {author} {\bibfnamefont {E.~A.}\ \bibnamefont
  {Quintero}}, \bibinfo {author} {\bibfnamefont {Charles~R.}\ \bibnamefont
  {Harris}}, \bibinfo {author} {\bibfnamefont {Anne~M.}\ \bibnamefont
  {Archibald}}, \bibinfo {author} {\bibfnamefont {Ant{\^o}nio~H.}\ \bibnamefont
  {Ribeiro}}, \bibinfo {author} {\bibfnamefont {Fabian}\ \bibnamefont
  {Pedregosa}}, \bibinfo {author} {\bibfnamefont {Paul}\ \bibnamefont {{van
  Mulbregt}}}, \ and\ \bibinfo {author} {\bibnamefont {{SciPy 1.0
  Contributors}}},\ }\bibfield  {title} {\enquote {\bibinfo {title} {{{SciPy}
  1.0: Fundamental Algorithms for Scientific Computing in Python}},}\ }\href
  {\doibase 10.1038/s41592-019-0686-2} {\bibfield  {journal} {\bibinfo
  {journal} {Nature Methods}\ }\textbf {\bibinfo {volume} {17}},\ \bibinfo
  {pages} {261--272} (\bibinfo {year} {2020})}\BibitemShut {NoStop}%
\bibitem [{\citenamefont {Rossi}\ \emph {et~al.}(2021)\citenamefont {Rossi},
  \citenamefont {Lu}, \citenamefont {Nag}, \citenamefont {Li}, \citenamefont
  {Osada}, \citenamefont {Lee}, \citenamefont {Wang}, \citenamefont
  {Agrestini}, \citenamefont {Garcia-Fernandez}, \citenamefont {Kas},
  \citenamefont {Chuang}, \citenamefont {Shen}, \citenamefont {Hwang},
  \citenamefont {Moritz}, \citenamefont {Zhou}, \citenamefont {Devereaux},\
  and\ \citenamefont {Lee}}]{rossi2020orbital}%
  \BibitemOpen
  \bibfield  {author} {\bibinfo {author} {\bibfnamefont {M.}~\bibnamefont
  {Rossi}}, \bibinfo {author} {\bibfnamefont {H.}~\bibnamefont {Lu}}, \bibinfo
  {author} {\bibfnamefont {A.}~\bibnamefont {Nag}}, \bibinfo {author}
  {\bibfnamefont {D.}~\bibnamefont {Li}}, \bibinfo {author} {\bibfnamefont
  {M.}~\bibnamefont {Osada}}, \bibinfo {author} {\bibfnamefont
  {K.}~\bibnamefont {Lee}}, \bibinfo {author} {\bibfnamefont {B.~Y.}\
  \bibnamefont {Wang}}, \bibinfo {author} {\bibfnamefont {S.}~\bibnamefont
  {Agrestini}}, \bibinfo {author} {\bibfnamefont {M.}~\bibnamefont
  {Garcia-Fernandez}}, \bibinfo {author} {\bibfnamefont {J.~J.}\ \bibnamefont
  {Kas}}, \bibinfo {author} {\bibfnamefont {Y.-D.}\ \bibnamefont {Chuang}},
  \bibinfo {author} {\bibfnamefont {Z.~X.}\ \bibnamefont {Shen}}, \bibinfo
  {author} {\bibfnamefont {H.~Y.}\ \bibnamefont {Hwang}}, \bibinfo {author}
  {\bibfnamefont {B.}~\bibnamefont {Moritz}}, \bibinfo {author} {\bibfnamefont
  {Ke-Jin}\ \bibnamefont {Zhou}}, \bibinfo {author} {\bibfnamefont {T.~P.}\
  \bibnamefont {Devereaux}}, \ and\ \bibinfo {author} {\bibfnamefont {W.~S.}\
  \bibnamefont {Lee}},\ }\bibfield  {title} {\enquote {\bibinfo {title}
  {Orbital and spin character of doped carriers in infinite-layer
  nickelates},}\ }\href {\doibase 10.1103/PhysRevB.104.L220505} {\bibfield
  {journal} {\bibinfo  {journal} {Physical Review B}\ }\textbf {\bibinfo
  {volume} {104}},\ \bibinfo {pages} {L220505} (\bibinfo {year}
  {2021})}\BibitemShut {NoStop}%
\bibitem [{\citenamefont {Hepting}\ \emph {et~al.}(2020)\citenamefont
  {Hepting}, \citenamefont {Li}, \citenamefont {Jia}, \citenamefont {Lu},
  \citenamefont {Paris}, \citenamefont {Tseng}, \citenamefont {Feng},
  \citenamefont {Osada}, \citenamefont {Been}, \citenamefont {Hikita} \emph
  {et~al.}}]{hepting2020electronic}%
  \BibitemOpen
  \bibfield  {author} {\bibinfo {author} {\bibfnamefont {Matthias}\
  \bibnamefont {Hepting}}, \bibinfo {author} {\bibfnamefont {Danfeng}\
  \bibnamefont {Li}}, \bibinfo {author} {\bibfnamefont {CJ}~\bibnamefont
  {Jia}}, \bibinfo {author} {\bibfnamefont {Haiyu}\ \bibnamefont {Lu}},
  \bibinfo {author} {\bibfnamefont {E}~\bibnamefont {Paris}}, \bibinfo {author}
  {\bibfnamefont {Y}~\bibnamefont {Tseng}}, \bibinfo {author} {\bibfnamefont
  {X}~\bibnamefont {Feng}}, \bibinfo {author} {\bibfnamefont {M}~\bibnamefont
  {Osada}}, \bibinfo {author} {\bibfnamefont {E}~\bibnamefont {Been}}, \bibinfo
  {author} {\bibfnamefont {Y}~\bibnamefont {Hikita}},  \emph {et~al.},\
  }\bibfield  {title} {\enquote {\bibinfo {title} {Electronic structure of the
  parent compound of superconducting infinite-layer nickelates},}\ }\href@noop
  {} {\bibfield  {journal} {\bibinfo  {journal} {Nature Materials}\ }\textbf
  {\bibinfo {volume} {19}},\ \bibinfo {pages} {381--385} (\bibinfo {year}
  {2020})}\BibitemShut {NoStop}%
\bibitem [{\citenamefont {{Arovas}}\ \emph {et~al.}(2021)\citenamefont
  {{Arovas}}, \citenamefont {{Berg}}, \citenamefont {{Kivelson}},\ and\
  \citenamefont {{Raghu}}}]{Arovas}%
  \BibitemOpen
  \bibfield  {author} {\bibinfo {author} {\bibfnamefont {Daniel~P.}\
  \bibnamefont {{Arovas}}}, \bibinfo {author} {\bibfnamefont {Erez}\
  \bibnamefont {{Berg}}}, \bibinfo {author} {\bibfnamefont {Steven}\
  \bibnamefont {{Kivelson}}}, \ and\ \bibinfo {author} {\bibfnamefont
  {Srinivas}\ \bibnamefont {{Raghu}}},\ }\bibfield  {title} {\enquote {\bibinfo
  {title} {{The Hubbard Model}},}\ }\href@noop {} {\bibfield  {journal}
  {\bibinfo  {journal} {arXiv e-prints}\ ,\ \bibinfo {eid} {arXiv:2103.12097}}
  (\bibinfo {year} {2021})},\ \Eprint {http://arxiv.org/abs/2103.12097}
  {arXiv:2103.12097 [cond-mat.str-el]} \BibitemShut {NoStop}%
\bibitem [{\citenamefont {Huang}\ \emph {et~al.}(2021)\citenamefont {Huang},
  \citenamefont {Wang}, \citenamefont {Ding}, \citenamefont {Liu},
  \citenamefont {Liu}, \citenamefont {Huang}, \citenamefont {Moritz},\ and\
  \citenamefont {Devereaux}}]{JPSJ}%
  \BibitemOpen
  \bibfield  {author} {\bibinfo {author} {\bibfnamefont {Edwin~W.}\
  \bibnamefont {Huang}}, \bibinfo {author} {\bibfnamefont {Wen~O.}\
  \bibnamefont {Wang}}, \bibinfo {author} {\bibfnamefont {Jixun~K.}\
  \bibnamefont {Ding}}, \bibinfo {author} {\bibfnamefont {Tianyi}\ \bibnamefont
  {Liu}}, \bibinfo {author} {\bibfnamefont {Fangze}\ \bibnamefont {Liu}},
  \bibinfo {author} {\bibfnamefont {Xu-Xin}\ \bibnamefont {Huang}}, \bibinfo
  {author} {\bibfnamefont {Brian}\ \bibnamefont {Moritz}}, \ and\ \bibinfo
  {author} {\bibfnamefont {Thomas~P.}\ \bibnamefont {Devereaux}},\ }\bibfield
  {title} {\enquote {\bibinfo {title} {{Intertwined States at Finite
  Temperatures in the Hubbard Model}},}\ }\href {\doibase
  10.7566/JPSJ.90.111010} {\bibfield  {journal} {\bibinfo  {journal} {Journal
  of the Physical Society of Japan}\ }\textbf {\bibinfo {volume} {90}},\
  \bibinfo {pages} {111010} (\bibinfo {year} {2021})},\ \Eprint
  {http://arxiv.org/abs/https://doi.org/10.7566/JPSJ.90.111010}
  {https://doi.org/10.7566/JPSJ.90.111010} \BibitemShut {NoStop}%
\bibitem [{\citenamefont {{Peng}}\ \emph {et~al.}(2021)\citenamefont {{Peng}},
  \citenamefont {{Jiang}}, \citenamefont {{Moritz}}, \citenamefont
  {{Devereaux}},\ and\ \citenamefont {{Jia}}}]{Peng_DMRG_arxiv}%
  \BibitemOpen
  \bibfield  {author} {\bibinfo {author} {\bibfnamefont {Cheng}\ \bibnamefont
  {{Peng}}}, \bibinfo {author} {\bibfnamefont {Hong-Chen}\ \bibnamefont
  {{Jiang}}}, \bibinfo {author} {\bibfnamefont {Brian}\ \bibnamefont
  {{Moritz}}}, \bibinfo {author} {\bibfnamefont {Thomas~P.}\ \bibnamefont
  {{Devereaux}}}, \ and\ \bibinfo {author} {\bibfnamefont {Chunjing}\
  \bibnamefont {{Jia}}},\ }\bibfield  {title} {\enquote {\bibinfo {title}
  {{Superconductivity in a minimal two-band model for infinite-layer
  nickelates}},}\ }\href@noop {} {\bibfield  {journal} {\bibinfo  {journal}
  {arXiv e-prints}\ ,\ \bibinfo {eid} {arXiv:2110.07593}} (\bibinfo {year}
  {2021})},\ \Eprint {http://arxiv.org/abs/2110.07593} {arXiv:2110.07593
  [cond-mat.str-el]} \BibitemShut {NoStop}%
\bibitem [{\citenamefont {{Liu}}\ and\ \citenamefont {{et
  al.}}(2021)}]{Fangze}%
  \BibitemOpen
  \bibfield  {author} {\bibinfo {author} {\bibfnamefont {Fangze}\ \bibnamefont
  {{Liu}}}\ and\ \bibinfo {author} {\bibnamefont {{et al.}}},\ }\href@noop {}
  {\bibfield  {journal} {\bibinfo  {journal} {unpublished}\ } (\bibinfo {year}
  {2021})}\BibitemShut {NoStop}%
\end{thebibliography}%

\end{document}